\documentclass{spie}
\usepackage[utf8]{inputenc}
\usepackage[letterpaper]{geometry}
\usepackage{graphicx}
\usepackage{aas_macros}
\usepackage{sidecap}

\title{The Cosmology Large Angular Scale Surveyor}
\author{Kathleen Harrington,\supit{a} Tobias Marriage,\supit{a}  Aamir Ali,\supit{a} John W. Appel,\supit{a} Charles L. Bennett,\supit{a} Fletcher Boone,\supit{e} Michael Brewer,\supit{a} Manwei Chan,\supit{a} David T. Chuss,\supit{d} Felipe Colazo,\supit{c} Sumit Dahal,\supit{a} Kevin Denis,\supit{c} Rolando Dünner,\supit{f} Joseph Eimer,\supit{a} Thomas Essinger-Hileman,\supit{a} Pedro Fluxa,\supit{f} Mark Halpern,\supit{b} Gene Hilton,\supit{g} Gary F. Hinshaw,\supit{b} Johannes Hubmayr,\supit{g} Jeffery Iuliano,\supit{a} John Karakla,\supit{a} Jeff McMahon,\supit{e} Nathan T. Miller,\supit{a,c} Samuel H. Moseley,\supit{c} Gonzalo Palma,\supit{i} Lucas Parker,\supit{a} Matthew Petroff,\supit{a} Bastián Pradenas,\supit{i} Karwan Rostem,\supit{a,c} Marco Sagliocca,\supit{d} Deniz Valle,\supit{a} Duncan Watts,\supit{a} Edward Wollack,\supit{c} Zhilei Xu,\supit{a} Lingzhen Zeng\supit{h}
\skiplinehalf
\supit{a} Dept. of Physics and Astronomy, Johns Hopkins University, Baltimore, MD, 21218, USA; \\
\supit{b} Dept. of Physics and Astronomy, University of British Columbia, Vancouver, BC V6T 1Z4, Canada; \\
\supit{c} Code 660, NASA Goddard Space Flight Center, Greenbelt, MD 20771, USA;\\
\supit{d} Dept. of Physics, Villanova University, Villanova, PA, 19085, USA;\\
\supit{e} Dept. of Physics, University of Michigan, Ann Arbor, MI, 48109, USA; \\
\supit{f} Instituto de Astrof\'isica and Centro de Astro-Ingenier\'ia,
Facultad de F\'isica, Pontificia Universidad Cat\'olica de Chile, Av.
Vicu\~na Mackenna 4860, 7820436 Macul, Santiago, Chile; \\
\supit{g} National Institute of Standards and Technology, 325 Broadway, Boulder, CO 80305, USA; \\
\supit{h} Harvard-Smithsonian Center for Astrophysics, Cambridge, MA 02138, USA; \\ 
\supit{i} Physics Department, FCFM, Universidad de Chile Blanco Encalada 2008, Santiago, Chile;
}

\authorinfo{Further author information: (Send correspondence to K. Harrington)\\K. Harrington.: E-mail: kharrington@jhu.edu}

\begin{document}

\maketitle

\section{Abstract}

The Cosmology Large Angular Scale Surveyor (CLASS) is a four telescope array designed to characterize relic primordial gravitational waves from inflation and the optical depth to reionization through a measurement of the polarized cosmic microwave background (CMB) on the largest angular scales.  The frequencies of the four CLASS telescopes, one at 38 GHz, two at 93 GHz, and one dichroic system at 145/217 GHz, are chosen to avoid spectral regions of high atmospheric emission and span the minimum of the polarized Galactic foregrounds: synchrotron emission at lower frequencies and dust emission at higher frequencies. Low-noise transition edge sensor detectors and a rapid front-end polarization modulator provide a unique combination of high sensitivity, stability, and control of systematics. The CLASS site, at 5200 m in the Chilean Atacama desert, allows for daily mapping of up to 70\% of the sky and enables the characterization of CMB polarization at the largest angular scales. Using this combination of a broad frequency range, large sky coverage, control over systematics, and high sensitivity, CLASS will observe the reionization and recombination peaks of the CMB E- and B-mode power spectra.  CLASS will make a cosmic variance limited measurement of the optical depth to reionization and will measure or place upper limits on the tensor-to-scalar ratio, $r$, down to a level of 0.01 (95\% C.L.).

\section{Introduction}

Measurements of the primordial temperature fluctuations in the Cosmic Microwave Background (CMB) have largely reached the cosmic variance limit \cite{WMAP9yrHinshaw,planck/15:2015}. These measurements have established the $\Lambda$CDM cosmological model and provided compelling evidence for Inflation. With the temperature data and their cosmological dividends nearly complete, attention has shifted to precision measurements of CMB polarization. Major goals include (1)  detection and characterization of inflationary gravitational waves through the divergence-free polarization pattern (``B-modes''), (2) improved estimates of the optical depth to reionization through the large-scale curl-free polarization pattern (``E-modes''), and (3) constraints on large scale structure and neutrino physics through gravitational-lensing-induced correlations within and between the CMB temperature fluctuations, E-modes, and B-modes. The challenges associated with these measurements are substantial. The polarization signals are orders of magnitude fainter than the temperature fluctuations and are obscured by polarized emission from the Milky Way.  To overcome these challenges and achieve the major science goals, a new generation of surveyors (so-called ``CMB Stage 3'') are pushing to new sensitivity limits with a range of experimental strategies.\cite{suzuki/etal:2016, ahmed/etal:2014, henderson/etal:2016,benson/etal:2014,ebex2010, PIPER2014, SPIDER2016, essinger/etal:2014}

\begin{figure}[t]
\centering
\includegraphics[width=0.35\textwidth]{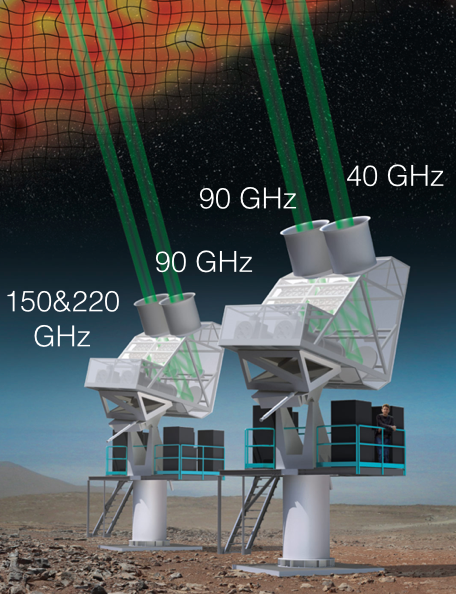}
\includegraphics[width=0.64\textwidth]{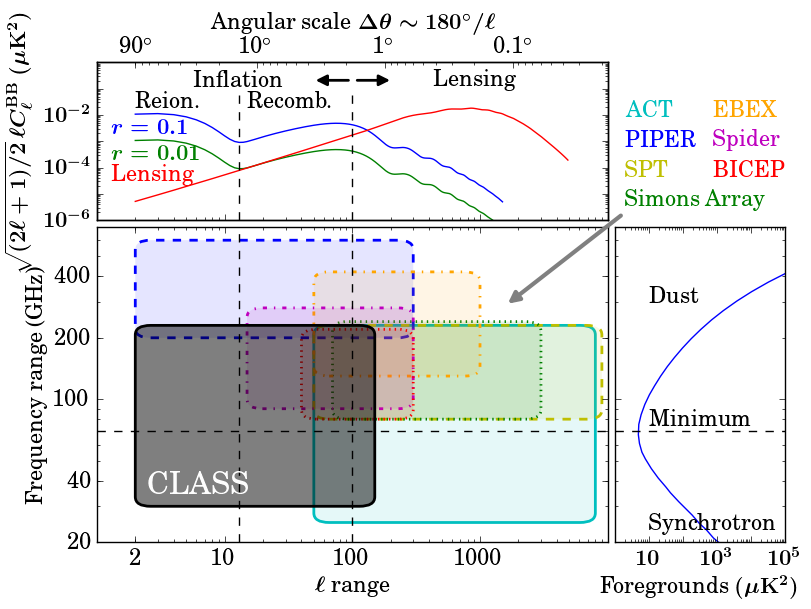}
\caption{\footnotesize {\bf The CLASS telescope array and survey.} The {\bf left figure} shows the four CLASS telescopes on two 3-axis  mounts rendered at the  5200~m site in the Atacama Desert. The telescopes operate across four frequencies straddling the Galactic foreground minimum. The polarization signal is rapidly modulated by the first optical element, enabling rejection of systematic effects and recovery of the large angular scale signal over nearly 70\% of the sky each day. The {\bf right figure} shows how the CLASS survey is uniquely designed to measure the primordial B-mode signal from both reionization and recombination over a frequency range straddling the foreground minimum. The figure gives the multipole ($\ell$) and frequency range of current surveys with forecasted constraints at the $r\sim0.01$ level, similar to CLASS. Top and side plots show the B-mode angular power spectrum  and the  frequency spectrum of polarized dust emission and synchrotron radiation.}
\label{fig:CLASS}
\end{figure}

Among the new generation of CMB polarimeters, the Cosmology Large Angular Scale Surveyor\cite{essinger/etal:2014} (CLASS; Figure \ref{fig:CLASS}) plays a unique and critical role. From 5200~m in the Atacama Desert of Chile, the CLASS  telescope array observes with two 93 GHz telescopes optimized for CMB observation near the minimum in the polarized Galactic emission, a 38 GHz telescope probes polarized synchrotron emission, and a 145/217 GHz telescope probes polarized dust. Instrument stability through front-end polarization modulation and a 200 sq-deg field-of-view enable recovery of CMB polarization at angular scales $\ge 2^\circ$ over nearly 70\% of the sky \textit{every day}. In this way CLASS measures over the full angular range of the predicted primordial B-mode signal. In particular CLASS is designed to be uniquely sensitive to the CMB polarization imprinted at reionization at angular scales of $\ge 10^\circ$. We describe the key role CLASS plays in realizing the major Inflation, reionization, and neutrino physics science goals. In this paper, we then provide an overview and status on the CLASS instrument and survey, focusing on the completed 38~GHz telescope. We refer the reader to previous CLASS publications for more details on the instrument, \cite{eimer/etal:2012,essinger/etal:2014,appel/etal:2014} simulations of polarized foreground treatments, \cite{watts/etal:2015} and systematic error rejection with the CLASS polarization modulator \cite{miller/etal:2016}.

\section{CLASS Science Outcomes}

\subsection{Inflation}

Inflation\cite{guth:1981,sato:1981,linde:1982,starobinsky:1982,albrecht/steinhardt:1982}, a rapid exponential expansion in the early universe that converts quantum to classical fluctuations \cite{mukhanov/chibisov:1981}, is foundational to the standard model of cosmology. The current observational imperative is to determine if the Inflation paradigm continues to hold and to better constrain the kind of inflation that could have happened. Inflation results in both scalar (density) perturbations with amplitude $A_s$ and tensor (gravitational wave) perturbations with amplitude $A_t$. Measurements of the CMB temperature and large-scale E-mode polarization provide the best constraints on $A_s$\cite{WMAP9yrHinshaw,planck/15:2015}, whereas B-mode measurements are required for progress on tensor perturbations. Therefore the highest priority of CMB polarization is to convincingly characterize primordial B-modes through a measurement of the tensor-to-scalar ratio $r$. 

\begin{figure}[t]
\centering
\includegraphics[width=0.52\textwidth]{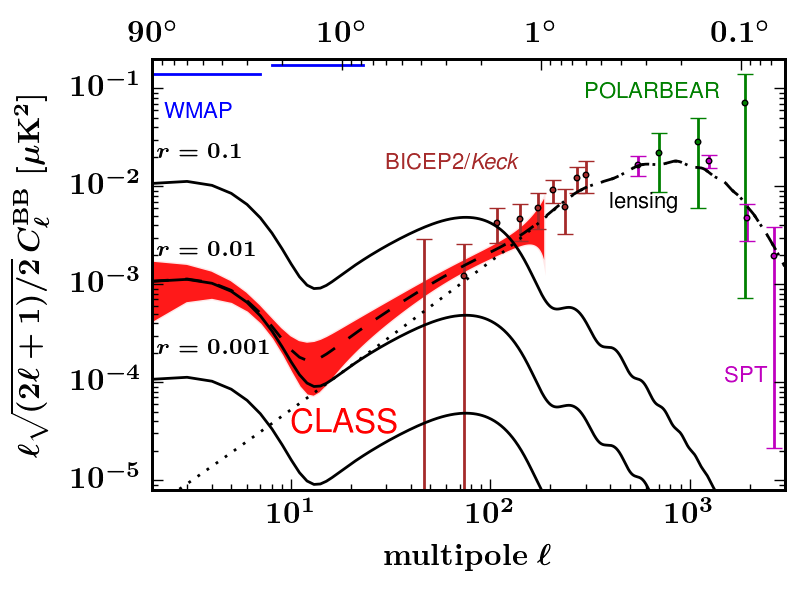}
\includegraphics[width=0.47\textwidth]{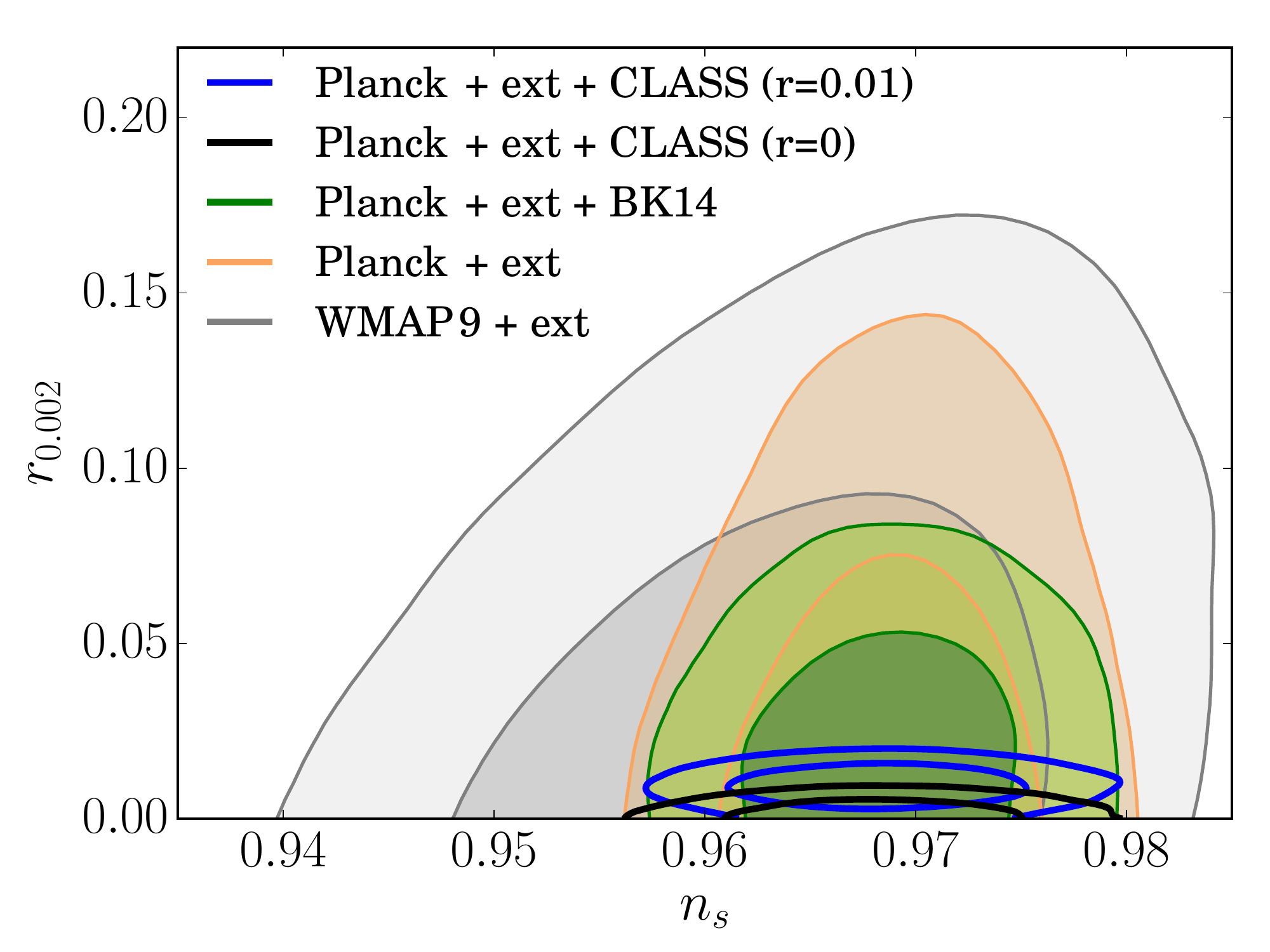}
\caption{\footnotesize \textbf{The CLASS B-mode Measurement.}  The {\bf left figure} shows power spectra for  B-mode amplitudes of $r=0.1$, $0.01$, and $0.001$ with  peaks at multipole moments $\ell < 10$ (the ``reionization peak'') and $\ell \approx 100$ (the ``recombination peak''). The dashed curve shows the  lensing of E-modes dominates the B-modes at small angular scales. The level was confirmed by SPT, P\textsc{olarbear},  and BICEP2/{\it Keck} \cite{keisler/etal:2015,polarbear/etal:2014,bk6}. The lensing B-modes have higher amplitude than the primordial signal for $r=0.01$ except at $\ell<20$, where the lensing foreground is negligible. With the exception of CLASS, all on-going measurements are designed to focus on the recombination peak and thus do not reach the largest scales (Figure \ref{fig:CLASS}). Due to its unique design CLASS probes all angular scales where the B-modes are strongest with sensitivity per $\log(\ell)$ shown in red. Multi-frequency measurements will be used to clean the Galactic foregrounds. {\it WMAP} measurements limit the B-mode amplitude at low-$\ell$ \cite{WMAP9yrBennett}. The {\bf right figure} shows upper limits on $r$ and measurements of $n_s$ as deduced from datasets primarily based on {\it WMAP} \cite{WMAP9yrHinshaw}, {\it Planck}  \cite{planck/15:2015}, and  BICEP2/{\it Keck} (BK14) with \textit{Planck}   \cite{bk6}. Forecasted CLASS constraints from simulations that include pessimistic effects of Galactic foregrounds are shown for two different simulated B-mode levels \cite{watts/etal:2015}.  CLASS will substantially reduce the upper limits on $r$ or determine $r$ in a reliable and robust manner. }
\label{fig:bmodes}
\end{figure}

Figure \ref{fig:bmodes} shows progress to date and forecasts for the CLASS B-mode measurement. The left plot shows that SPT, Polarbear, and BICEP2/{\it Keck} have detected B-mode polarization due to gravitational lensing of E-mode polarization\cite{keisler/etal:2015,polarbear/etal:2014,bk6}. The BICEP2/{\it Keck} polarization measurements at $\ell \sim 100$ offer the most stringent polarization-only constraint on $r$ to date with  $r<0.09$ (95\% C.L.) \cite{bk6}. Constraints improve to $r<0.07$ (95\% C.L.) if {\it Planck} temperature data are included \cite{bk6}. The measurements at $\ell\sim100$, the ``recombination peak'', are susceptible to the lensing foreground whereas this contamination is negligible for the ``reionization peak'' at $\ell<10$. The CLASS sensitivity per $\log(\ell)$ for $r=0.01$ is shown in the left plot by the red envelope, indicating sensitivity to both the recombination and reionization peaks. The right plot of Figure \ref{fig:bmodes} shows that CLASS will provide roughly an order-of-magnitude improvement in raw sensitivity to $r$ \cite{watts/etal:2015, miller2016}. Measurement of a double-peaked B-mode spectrum is an important confirmation that a systematic error is not masquerading as the faint primordial B-mode signal, especially if the recombination peak is significantly below the level of lensing B-modes.

\subsection{Reionization}

\begin{figure}[t]
\centering
\includegraphics[width=0.57\textwidth]{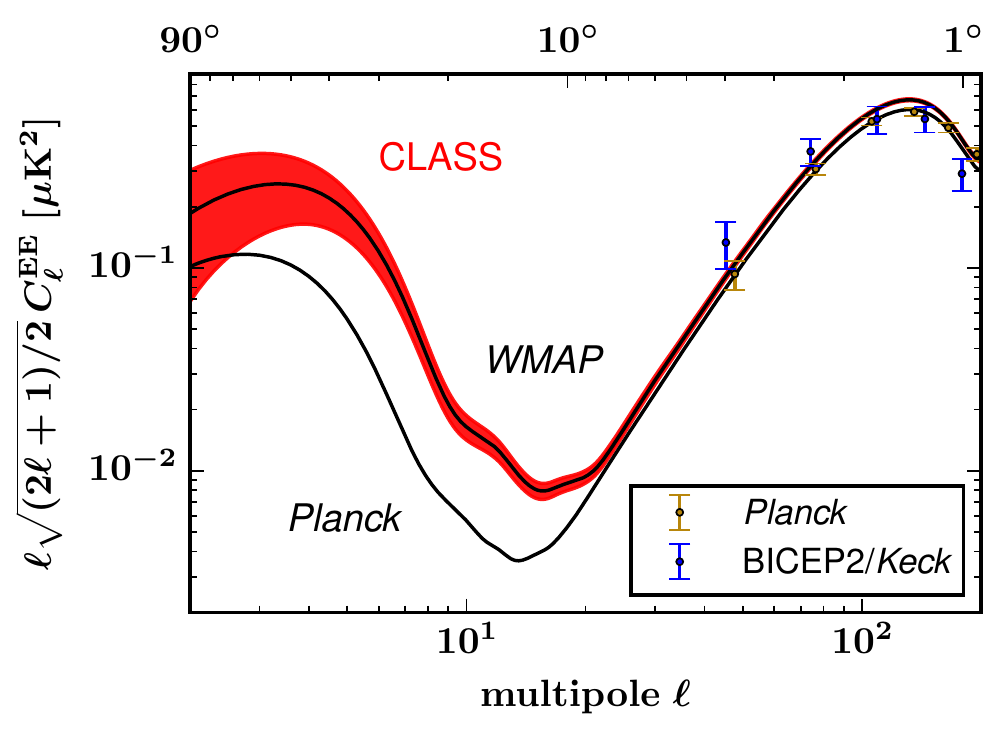}
\includegraphics[width=0.42\textwidth]{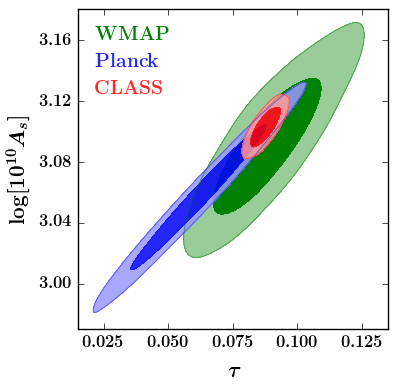}
\caption{\footnotesize {\bf Large-scale E-mode Science.} The {\bf left figure} shows  measurements and best-fit models for the low-$\ell$ E-mode polarization spectrum. The model for the upper black curve derives from WMAP 9-year (+SPT+ACT+SNLS3+BAO+H0) data, corresponding to $\tau=0.083$, whereas the model for the lower black curve shows the {\it Planck} ``pre-2016 lowE'' result, corresponding to $\tau=0.055$ \cite{planck/intermediate/XLVI}. The red envelope, which represents the CLASS survey  noise per $\log(\ell)$, shows that CLASS will discriminate between the {\it WMAP} and {\it Planck} best fit models, a result with implications for high-$z$ galaxy formation. The data at $\ell \ge 50$ are the latest from the {\it Planck} HFI and BICEP2/{\it Keck}. CLASS is stable enough to reach lower $\ell$ and complete the E-mode measurement. The {\bf right figure} shows constraints on $\tau$ and $\log(10^{10}A_s)$ from {\it WMAP} (green) and {\it Planck} (blue) temperature and polarization \cite{WMAP9yrHinshaw,planck2015XIII}.
 The {\it Planck} contours are based on the 2015 release instead of the more recent ``pre-2016'' results\cite{planck/intermediate/XLVI}
 because likelihoods were not released for the latter. The uncertainty is dominated by degeneracy between the two parameters.  The CLASS $\tau$ measurement breaks this degeneracy. (The WMAP9 value for $\tau$ has been assumed in the CLASS constraint.) }
\label{fig:emodes}
\end{figure} 

Improved measurements by CLASS of the E-mode polarization at the largest angular scales will help to further constrain the epoch of reionization, a major focus of current cosmology study. The CLASS survey will make a cosmic variance limited measurement of the E-mode spectrum below $\ell \approx 100$ (Figure \ref{fig:emodes}, left panel). The corresponding error on the optical depth $\tau$ to reionization is $\sigma_\tau\approx0.004$, an improvement by more than two over the  uncertainty of the ``pre-2016'' {\it Planck} results with ($\sigma_\tau=0.009$) \cite{planck/intermediate/XLVI}.  Furthermore given cosmic variance in the E-mode signal, the next generation polarization experiments (``Stage 4'' or a putative space mission) will not improve the CLASS E-mode measurement significantly.  This measurement will be an important complement and crosscheck to the current and next generation of 21-cm measurements of the epoch of reionization \cite{clesse/etal:2012, liu/etal:2015}.
Given its sensitivity to $\tau$, the CLASS survey is in a unique position to shed light on the tension that has arisen between the  {\it Planck}  and  {\it WMAP} estimates of the optical depth to reionization. The best fit {\it WMAP} 9-year (+SPT+ACT+SNLS3+BAO+H0) result is $\tau=0.083\pm0.013$ whereas the {\it Planck} pre-2016 (+BAO+JLA+H0) result is $\tau=0.055\pm0.009$. As shown in the left panel of Figure \ref{fig:emodes}, CLASS will easily discriminate between these two scenarios. 

\subsection{Neutrino Physics}

\begin{figure}[t]
    \centering
    \includegraphics[width=2.7in]{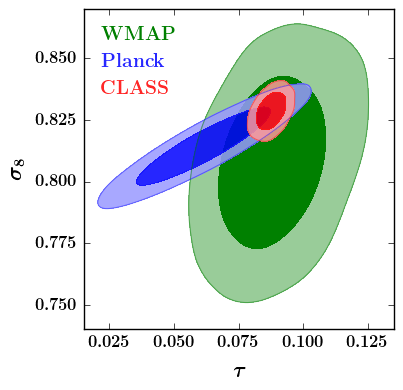}
    \includegraphics[width=2.7in]{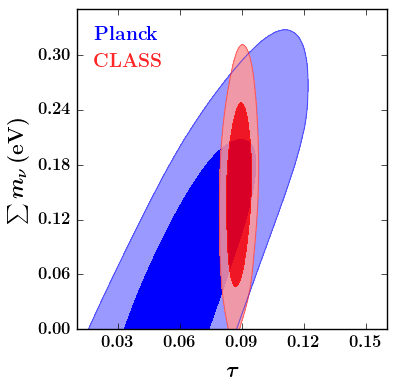}
    \caption{\footnotesize {\bf Improving CMB-based ${\Lambda}$CDM Predictions for LSS.} The $\Lambda$CDM model fit to the primary CMB provides predictions for LSS that can be compared to direct probes, such as galaxy weak lensing, galaxy cluster counts, and CMB lensing. However the CMB-based LSS predictions are now strongly limited by degeneracies, especially between $A_s$ and $\tau$ (Figure \ref{fig:emodes}). The {\bf left figure} shows how the CLASS E-mode measurement will improve predictions for $\sigma_8$. Reconciling the CMB LSS predictions with those from direct LSS probes   will provide tight constraints on the sum of neutrino masses $\sum m_\nu$. The {\bf right figure} shows improvement to the {\it Planck} 2015 and BAO \cite{planck2015XIII} constraints on $\sum m_\nu$. (The {\it Planck} contours are based on the 2015 data instead of the more recent ``pre-2016'' results since likelihoods are unavailable for the latter.) }
    \label{fig:sigma8}
\end{figure}

Beyond the direct application to studies of the epoch of reionization, the CLASS E-mode measurement will be crucial for studies of the evolution of large scale structure (LSS).  The $\Lambda$CDM model, as constrained by the CMB, gives predictions about the LSS today that can be compared to direct probes, such as galaxy surveys, galaxy cluster counts, or CMB lensing. Discrepancies between the CMB LSS prediction and the direct probes can be used to explore physics beyond the vanilla $\Lambda$CDM model. However, the CMB constraints on LSS are based on $A_s$ and are therefore limited by a degeneracy between the $A_s$ and $\tau$. CMB temperature fluctuations, which  increase with $A_s$, are suppressed by reionization such that the amplitude of the temperature spectrum constrains the parameter combination $A_s e^{-2\tau}$. This $A_s$ vs $\tau$ degeneracy is apparent in the right panel of Figure \ref{fig:emodes}. The uncertainty on $A_s$ is dominated by this degeneracy, so the CLASS $\tau$ constraint reduces the {\it Planck} 2015 or {\it WMAP}-based uncertainty on $\log(10^{10}A_s)$ by a factor of three (factor of 2 for {\it Planck} ``pre-2016''\cite{planck/intermediate/XLVI}). Figure \ref{fig:sigma8} shows the corresponding 2.5 times improvement in precision over the {\it Planck} 2015 estimate of $\sigma_8$ (two-fold improvement relative to {\it Planck} ``pre-2016''). Such an improvement will sharpen or resolve discrepancies arising between the predictions of the CMB-based $\Lambda$CDM models and direct probes of LSS, such as galaxy cluster measurements \cite{planck2015XXIV}. Such discrepancies can be due to systematic errors in measurements, and the CLASS $\tau$ measurement will be instrumental in removing these systematics. Beyond systematics, there \emph{are} physical effects, such as the finite neutrino mass, that should produce discrepancies that are detectable with the sensitivity of the next generation of CMB lensing (e.g., by Advanced ACTPol, Simons Array, and SPT3G or ``Stage 4'' surveys) and Baryon Acoustic Oscillation (BAO) measurements (e.g., DESI).\footnote{Other probes of LSS associated with galaxy and galaxy cluster surveys, such as  cluster counts and the matter power spectrum, will also constrain $\sum m_\nu$.  CMB lensing with BAO has the advantage of being relatively immune to systematics associated with baryonic processes.} However, these measurements will be limited by our uncertainty on $A_s$ without an improved $\tau$ measurement. Neutrino oscillation measurements, which are sensitive to the difference of the neutrino masses, put a lower bound of $\sum m_\nu \ge 60$~meV on the neutrino mass (100~meV for an inverted mass hierarchy with two heavy neutrinos) \cite{feldman/hartnell/kobayashi:2013}. The authors of Reference~\citenum{allison/etal:2015} investigate the power of the primary CMB in combination with the next generation of CMB lensing and BAO  measurements to constrain  $\sum m_\nu$. They show that without an improvement in the current low-$\ell$ E-mode measurement, the future experiments will be limited to $\sigma(\sum m_\nu) \approx 27$~meV (22~meV if {\it Planck} reaches it's ultimate low-$\ell$ sensitivity), which would provide only 2$\sigma$ evidence if $\sum m_\nu = 60$~meV. With a cosmic variance limited E-mode measurement, the parameter-covariance-dominated error reduces to $\sigma(\sum m_\nu) \approx 15$~meV, giving a 4$\sigma$ detection of the minimum-allowed neutrino mass. Thus the CLASS E-mode measurement is a key ingredient in pursuit of a cosmological measurement of the sum of neutrino masses.

\subsection{Further Science}

\textbf{Large-Scale Anomalies.} CMB anisotropy measurements covering all or most of the sky probe the largest accessible scales in the early universe. {\it WMAP} and {\it Planck} temperature data contain hints of several possible large-scale anomalies, including a deficit of power compared to $\Lambda$CDM predictions at low multipoles, an unusually large hemispherical asymmetry, and apparent dipole modulation of the power spectrum \cite{planck2015XVI}. While not of decisive significance, these observations have generated substantial interest in the community \cite{bennett/etal:2011,schwarz/etal:2015}. Physical models for such anomalies generically predict corresponding signatures in E-mode polarization. Measuring such signatures would point to a breakdown of statistical isotropy and have profound implications for our understanding of cosmology. Some tests were performed using {\it Planck} 2015 polarization data, however a mismatch of observed and simulated noise levels hampered a full investigation \cite{planck2015XVI}. CLASS will provide large-scale E-mode polarization measurements that are a factor of $5-10$ times more sensitive than {\it Planck} and allow the most stringent test to date of a range of models for large-scale anomalies.

\noindent \textbf{New Particles \& Circular Polarization.} The CMB travels to us over 13.8 billion years from the edge of the observable universe. During this long journey, subtle processes, previously undetected, can alter the CMB polarization in measurable ways. Grand Unified Theories predict the generation of new quantum fields in the early universe that could interact with the CMB (e.g., by a Chern-Simons mechanism) to rotate its linear polarization, an effect that CLASS is well suited to measure \cite{caldwell/etal:2011,zhao/etal:2014}.  CLASS is also uniquely poised to improve the precision of circular polarization measurements by two orders of magnitude \cite{mainini/etal:2013}. While circular polarization is predicted by theory to be absent in the CMB, the theory must be tested. In these ways the CLASS searches for new fundamental interactions and pushes the boundaries of physical theory.

\section{CLASS Instrument and Survey Design Overview} 
\label{sec:class}

For completeness, this section provides a summary of the CLASS instrument design and survey strategy. For more detailed descriptions see References \citenum{essinger/etal:2014,eimer/etal:2012,watts/etal:2015,miller2016}.

\subsection{The Instrument} 

CLASS measures the CMB polarization from 5200~m in the Atacama Desert of northern Chile. The site allows observation through low atmospheric water vapor and oxygen column densities, which translates to reduced atmospheric brightness in the CLASS frequency bands.
 
Figure \ref{fig:CLASS} shows the four CLASS telescopes with observing frequencies 38 GHz, 2$\times$93~GHz, and 145/217~GHz. The four telescopes share two telescope mounts. Each telescope mount rotates in three axes (azimuth, elevation, and boresight). The boresight rotation is a key ingredient for rejecting systematic effects. Each telescope has an approximately 200~deg$^2$ field of view and resolutions ranging from 12$'$ (217~GHz) to 90$'$ (38~GHz). See Reference~\citenum{essinger/etal:2014} for a full specification of frequency bandwidths and angular resolutions and Reference~\citenum{eimer/etal:2012} for a description of the optical design.

\begin{figure}[t]
\centering
\includegraphics[width=.9\textwidth]{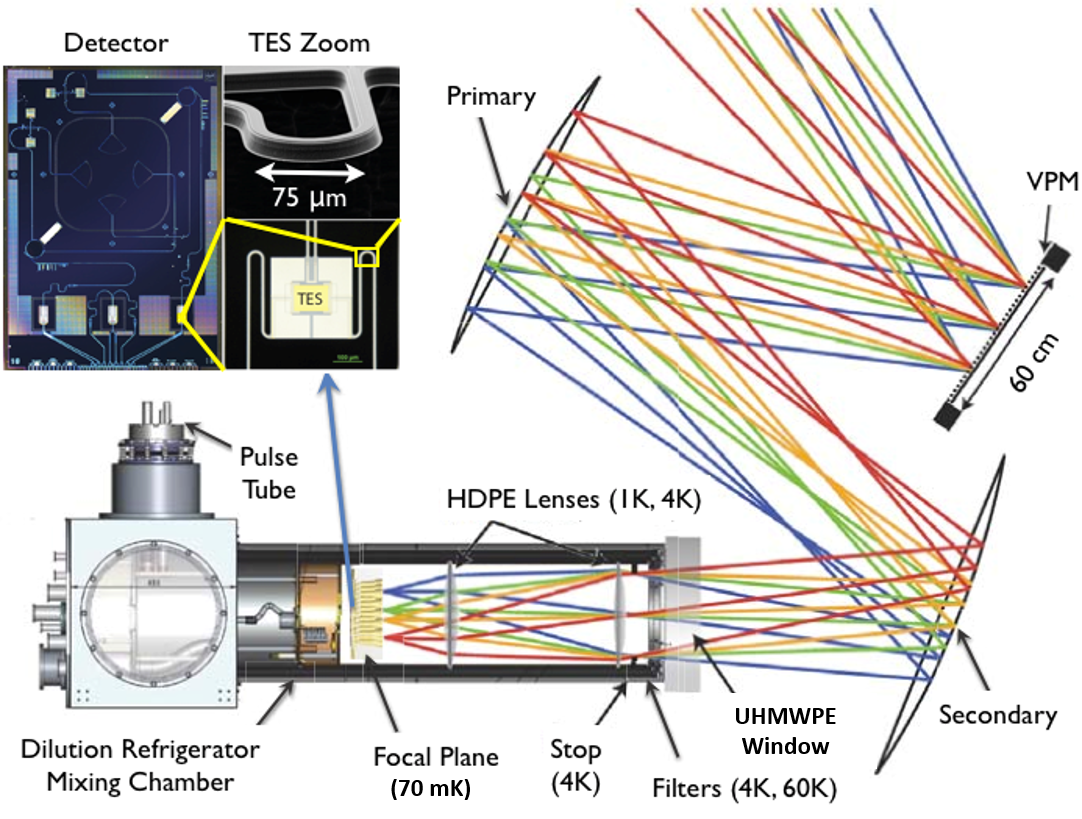}
\caption{\footnotesize {\bf A CLASS Telescope.} Each CLASS telescope is designed to minimize systematic effects while optimizing sensitivity.  The figure shows a schematic with colored lines tracing the optical path for four beams. The VPM modulates the polarized sky signal (e.g., the CMB) and, \emph{because it is the first optical element}, does not modulate instrumental polarization. Because atmospheric emission is not linearly polarized, it does not contaminate the signal. Mirrors and lenses focus the light onto the focal plane, which is continuously cooled to 70~mK in the dilution refrigerator-based cryogenic receiver. A highly-efficient polarization-sensitive detector defines the band-pass and measures the power of the linear polarization states with  transition edge sensor (TES) bolometers.}
\label{fig:telescope}
\end{figure}

The CLASS telescopes share a common novel design shown in  Figure \ref{fig:telescope} \cite{eimer/etal:2012, essinger/etal:2014}. The first, and most crucial, optical element is the variable-delay polarization modulator \cite{chuss/etal:2012,miller/etal:2016}. CLASS uses polarization modulation to move the signal band to 10 Hz, well above the $1/f$ knee of the instrument and environment. While observations without  polarization modulation and/or without a large ($>10\deg$) field of view may target B-modes from recombination at $\ell=100$, {\it the combination of the ``lock-in'' style stability afforded by the VPM together with CLASS's large FOV for simultaneously capturing the polarized sky over 200~deg$^2$ is a requirement to measure polarization on the largest angular scales.} A primary advantage of a VPM over other modulation technologies is that it straightforwardly scales to the large ($\sim60$~cm) size required for the first optical element: the VPM is placed as the \emph{first} optical element  to avoid contamination of the CMB polarization by instrumental polarization (the conversion of unpolarized light into polarized light by the telescope).

The VPM is followed by elliptical primary and secondary mirrors oversized to prevent warm spillover and to image the 4~K cold stop onto the VPM with low edge illumination \cite{eimer/etal:2012}.  Locating the VPM at a pupil ensures that each beam shares the same area of the modulator and thus common modulator response, while the reduced edge illumination eliminates systematics due to edge effects, including diffraction. 
After the mirrors, the light enters the cryostat where infrared radiation is rejected by filters and the remaining signal is imaged  onto the focal plane, which consists of an array of feedhorn-coupled, polarization-sensitive detectors continuously cooled to 70~mK by a dilution refrigerator. Feedhorns and on-chip detector circuitry define polarized beams that cleanly propagate through the telescope with high optical efficiency  over a large bandwidth. Two transition edge sensors (TESs) detect the power corresponding to orthogonal linear polarizations. The TESs are read out with time-domain Superconducting Quantum Interference Device multiplexing electronics \cite{irwin/etal:2002}. The CLASS  detectors are unique in the field of CMB polarimetry in their combination of wide bandwidth, reproducibility, high efficiency, out-of-band signal rejection, and overall control of systematics \cite{Denis/etal:2009,chuss/etal:2012a,rostem/etal:2014b,appel/etal:2014}.

\subsection{The Survey} 
\label{subsec:survey}

To measure CMB polarization on the largest angular scales, CLASS must survey a majority of the sky. By scanning the sky in large circles at a constant  45${^\circ}$ above the horizon, CLASS covers 70\% of the sky (-78${^\circ}$ ${< \delta <} $+32${^\circ}$).\footnote{This fraction takes into account CLASS's large 200~deg$^2$ field of view.} The scans cross-link one another so that each point on the sky is observed with the telescope oriented and scanning at different angles. Such cross-linking has been shown by {\it WMAP}, QUIET, ACT, and other surveys to be highly advantageous in rejecting systematic errors and reconstructing the large angular scale signals in the map-making process. In addition to scan cross-linking, the boresight angle of the observations is rotated daily by 15$^\circ$ to provide further systematics checks and to  break degeneracies when solving for the polarization maps from the raw bolometer time streams. CLASS will observe 24 hours per day, covering nearly the full survey area daily (allowing for sun-avoidance). With this strategy, we will diagnose survey performance and check for  systematic effects  on timescales  from days to years.  Emission from the VPM  produces a specific signature, uncorrelated with the polarized sky signal, that provides absolute calibration for the VPM encoders. Observations of sources (e.g., the moon, planets, Tau A) will be used to  measure telescope pointing, beams, and polarization angles. Ultimate calibration will take advantage of high-precision space mission calibration, correlating CLASS data with the {\it WMAP} and {\it Planck} polarization data (including sources like Tau~A and fits to data across the Galactic Plane). The CLASS bands are close to the {\it WMAP} and {\it Planck} bands, so Galactic emission spectral uncertainties do not propagate significantly to calibration uncertainties.   Additional tests, such as using an emissive sparse wire grid to measure relative detector angles will serve as checks of our primary on-sky calibration.

The total survey time is five years (2016--2021) with a staged deployment of the CLASS telescopes over the first two years. The 38~GHz telescope is now in the field and will be followed by the two 93~GHz telescopes and finally the 145/217~GHz telescope. For more information on integration times and survey sensitivies, see Reference~\citenum{essinger/etal:2014}. For consideration of instrument stability through rapid VPM modulation, see Reference~\citenum{miller/etal:2016}. For demonstration of Galactic foreground treatment for the CLASS survey, see Reference~\citenum{watts/etal:2015}.

\section{Instrument Update}

This section provides updates to the CLASS instrument since Reference~\citenum{essinger/etal:2014}. These updates primarily focus on the 38~GHz CLASS telescope, which has been deployed to the newly operational CLASS site in the Atacama Desert in Chile.

\subsection{Operations}
\begin{figure}[t]
    \centering
    \includegraphics[width=3.9in]{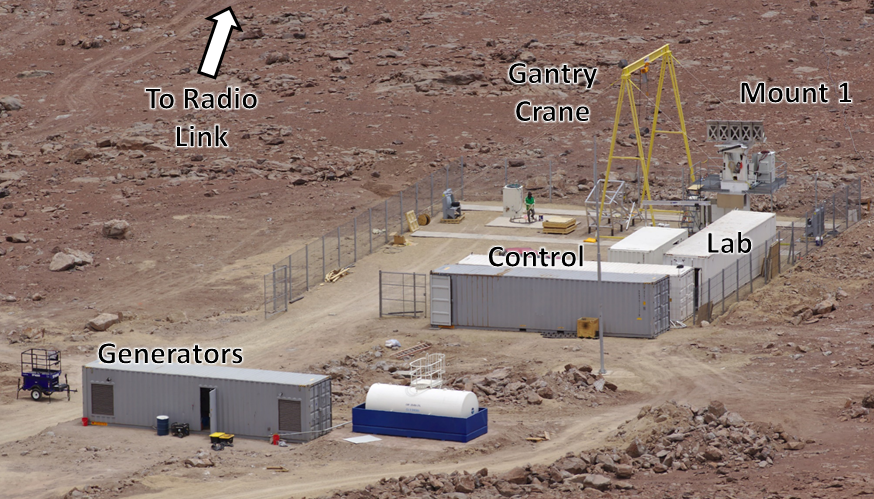}
    \includegraphics[width=1.55in]{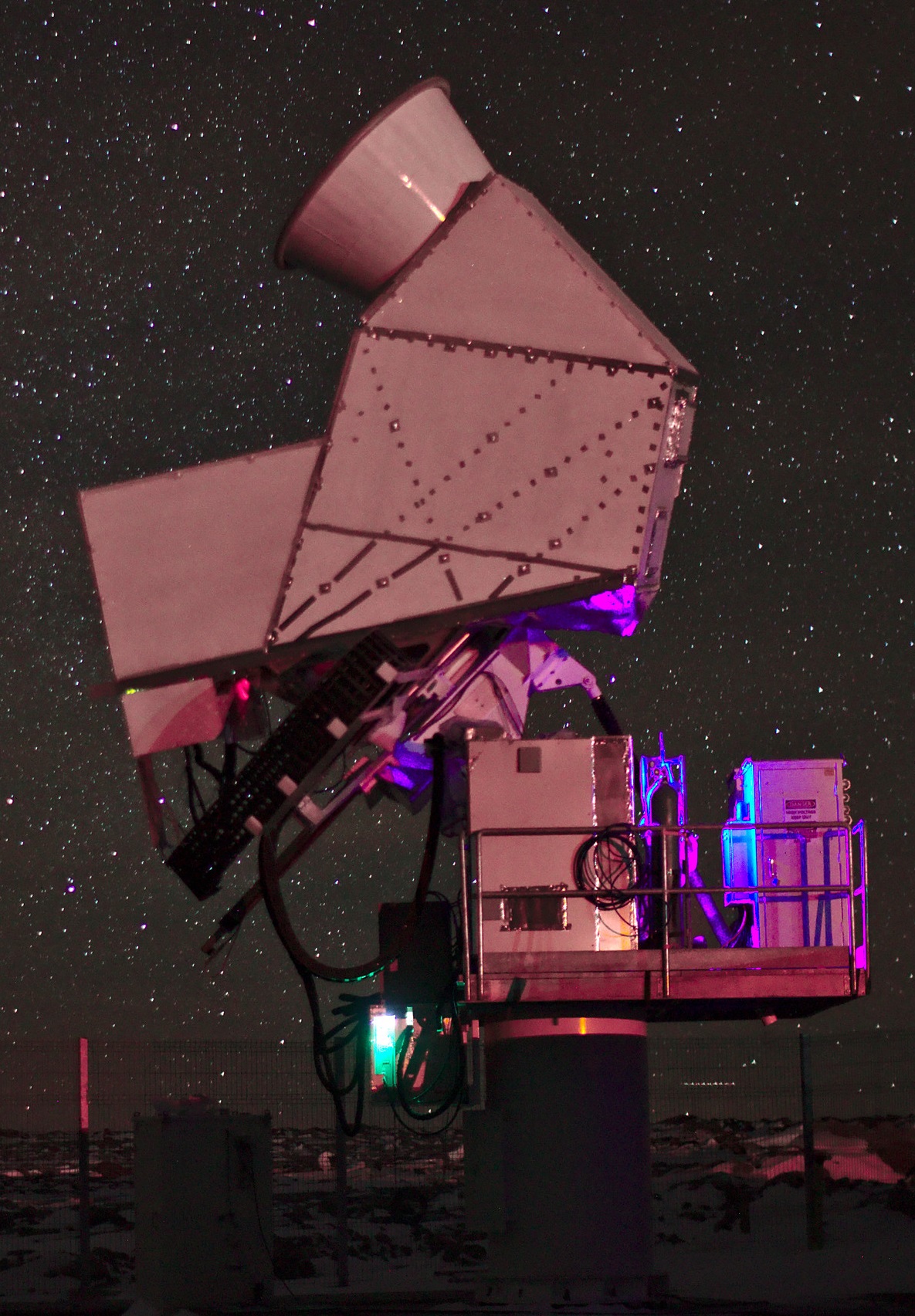}
    \caption{\footnotesize {\bf Left Figure:} The CLASS site as it was in January 2016. The generator container is in the foreground while the control, laboratory, and machine shop containers are inside the fenced area. The bare mount is shown before the receiver, telescope optics, and co-moving ground screen were installed. The yellow A-frame gantry crane was used to install the telescope components onto the mount. {\bf Right Figure:} The 38 GHz CLASS telescope installed at the CLASS site. Visible in the photograph are the co-moving ground screen, forebaffle, and platform containing the gas handling system for the receiver and the mount drive system.}
    \label{fig:class_site}
\end{figure}

The CLASS site, shown in Figure \ref{fig:class_site}, provides access to 70\% of the sky with minimal loading due to atmospheric water and oxygen.\cite{essinger/etal:2014} Construction on the site began in early 2015. The flattened, fenced site contains two concrete pedestals for the two CLASS mounts and a concrete `sidewalk' for the A-frame gantry crane used to assemble each telescope. Power is provided by two redundant diesel powered generators, and a radio link to the CLASS base of operations in San Pedro de Atacama allows communication with the site. A custom data packaging system transmits the telescope data via the radio link to San Pedro, where it is stored. The 38~GHz telescope data is transferred via the Internet to North America; future telescope deployments will produce higher data rates, and data transfer to North America will proceed through different methods.

The control room, laboratory, and machine shop containers, shipped with the first CLASS mount and 38~GHz receiver packed inside, arrived at the CLASS site in December 2015. The telescope mount was installed in January 2016 and the 38~GHz receiver was installed in the spring of 2016. The cryogenics are stable and able to hold all stages at their respective temperatures for the entire range of planned boresight rotations ($-45^{\circ} \leq \delta \leq 45^{\circ}$). 

The CLASS 38 GHz telescope achieved first light on May 8, 2016 and began a commissioning phase. Detector loading and saturation power are within their designed ranges. As of these proceedings, the 38 GHz telescope is performing nightly sky observations using the nominal CLASS observation strategy: constant 45$^\circ$ elevation scans at 1$^\circ/s$ across 360$^\circ$ in azimuth with the boresight angle stepped 15$^\circ$ per night.

\subsection{Mounts}	
The four CLASS telescopes share two telescope mounts that each have three axes of rotation: azimuth, elevation, and boresight. The azimuth axis has a maximum slew rate of $3^\circ/s$ and a range from $-200^\circ$ to $+560^\circ$ while the elevation axis has a maximum slew rate of $1^\circ/s$ and a range from $20^\circ$ to $90^\circ$ and the boresight axis has a maximum slew rate of $0.5^\circ/s$ and a range from $-45^\circ$ to $45^\circ$. Each axis has an RMS pointing accuracy of 15 arcseconds. A co-moving ground screen and 1.5 meter diameter forebaffle are used to mitigate ground pickup.

Prior to being shipped to the CLASS site, the first mount was fully assembled in the highbay at Johns Hopkins University (JHU) Department of Physics and Astronomy. During the summer of 2015 an all-systems integration test was performed at JHU. The 38 GHz receiver, optics, and VPM were installed on the mount; the receiver was cooled and the detectors were tested while the VPM was running and the mount was scanning. The second CLASS mount and additional receivers will follow the same path.

\subsection{Detectors}

CLASS uses feedhorn-coupled polarization sensitive detectors fabricated at the NASA Goddard Space Fight Center. A smooth-walled feedhorn and planar orthomode-transducer (OMT) couple two orthogonal polarizations onto planar microstrip transmission lines on single crystal silicon. An integrated quarter-wavelength backshort assembly completes the OMT and protects against stray light leakage. The monocrystalline silicon provides an extremely low loss dielectric for the microwave circuitry. Detection bands are defined by on-chip filters, after which the powers from two linear polarizations are terminated onto two Mo-Au transition edge sensors (TESs) with a transition temperature of $\sim150$~mK. The CLASS detector architecture is unique for its scalability, bandwidth, and noise properties. More details about the CLASS detectors can be found in Reference \citenum{appel/etal:2014} and \citenum{Rostem2016}.

\begin{figure}[t]
    \centering
    \includegraphics[width=2.1in]{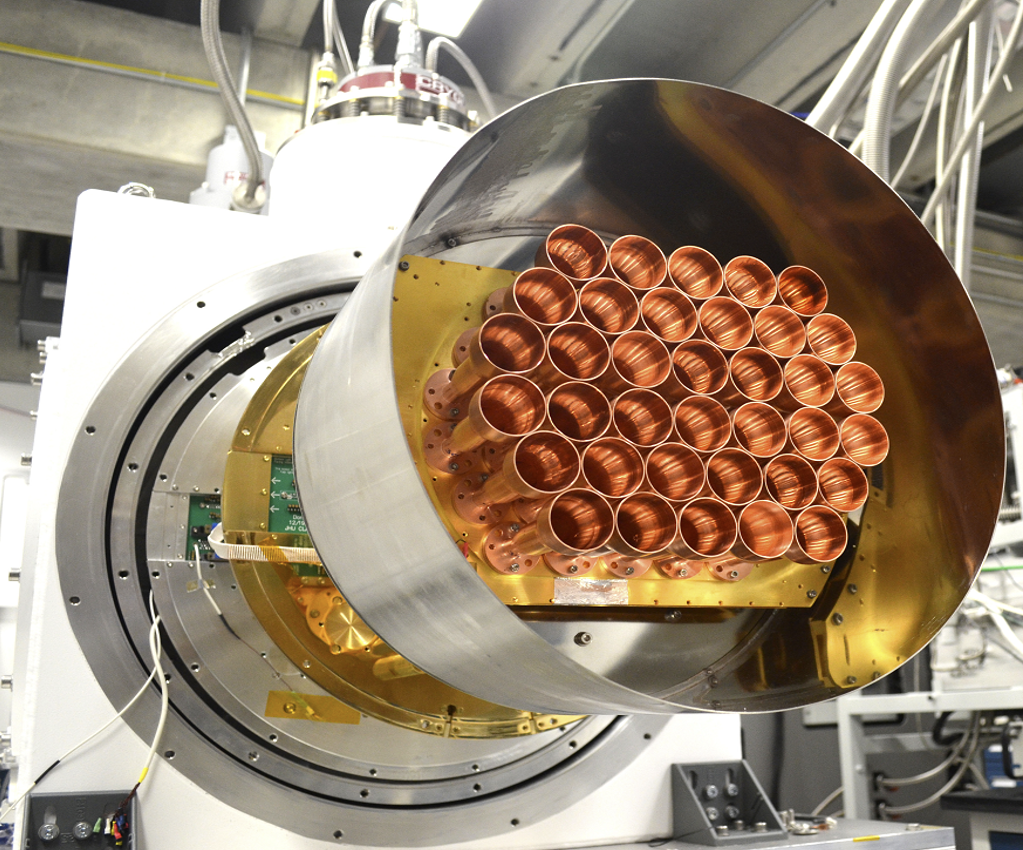}
    \includegraphics[width=3.3in]{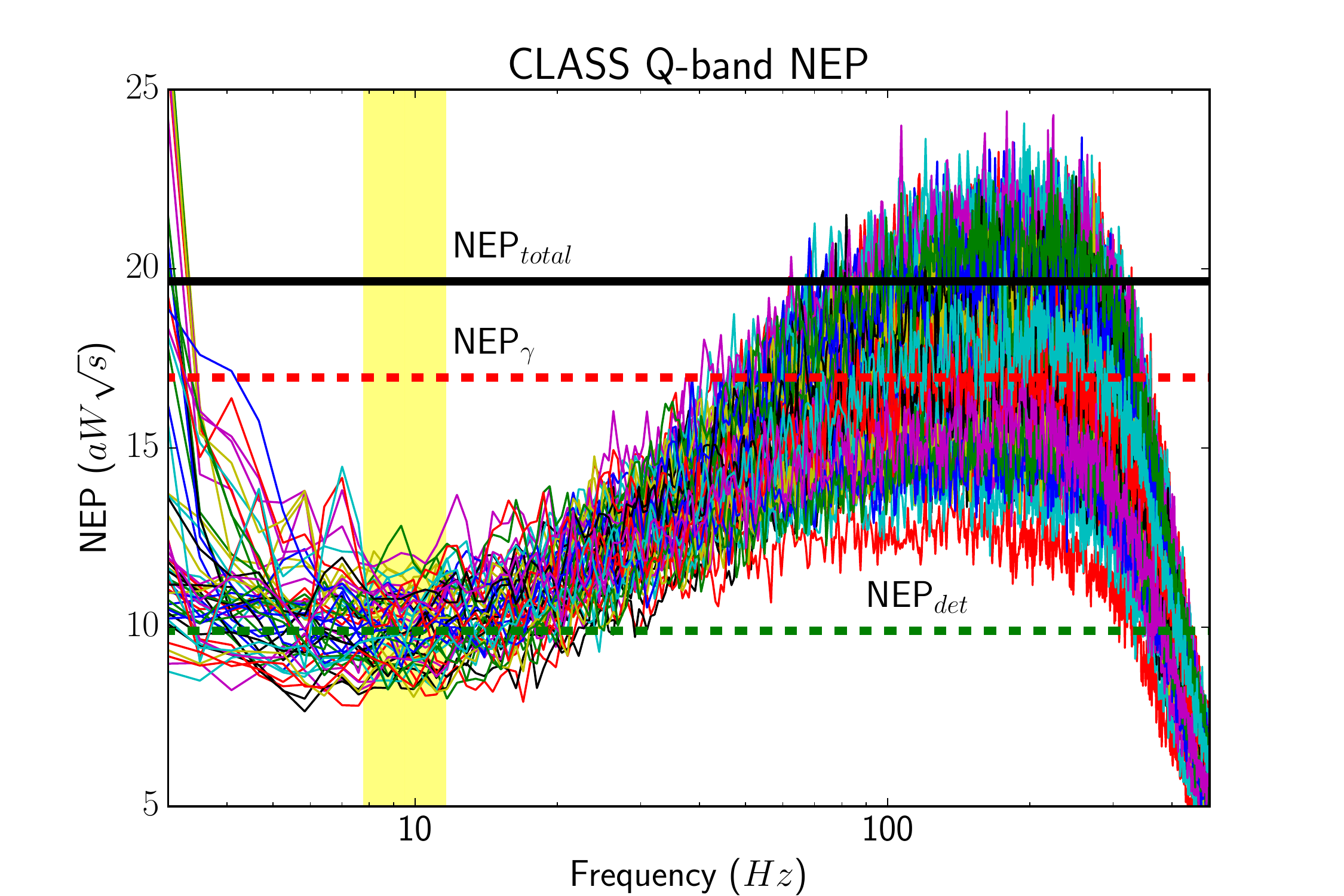}
    \caption{\footnotesize {\bf Left Figure:} The fully integrated CLASS 38 GHz focal plane mounted in the cryostat receiver. {\bf Right Figure:} The dark noise of the CLASS 38 GHz detectors. The average detector noise around the 10 Hz modulation frequency (green dashed line) is significantly lower than the expected noise due to optical loading on the detectors (red dashed line).}
    \label{fig:focal_planes}
\end{figure}

The CLASS 38 GHz focal plane includes 36 dual-polarization detectors mounted into a gold-plated copper baseplate. The left image in Figure~\ref{fig:focal_planes} shows the completed 38 GHz focal plane mounted inside the 38 GHz receiver. Laboratory measurements of dark detector noise are plotted in the right frame of Figure \ref{fig:focal_planes}. The average dark detector noise around the 10 Hz modulation frequency (green dashed line) is significantly lower than the expected photon noise contribution (red dashed line) meaning each detector is background limited.

The 93 GHz focal planes consist of seven hexagonal monolithic wafer modules with 37 dual-polarization detectors in each module. All modules for the first 93 GHz focal plane are in-hand for integration into a complete focal plane.

\subsection{Optics and Receivers}

Each telescope in the CLASS array has a similar optical design beginning with a variable-delay polarization modulator (VPM) as the first element from the sky. Two 1~m aluminum mirrors focus the light into the receiver where it is filtered to remove out-of-band power and imaged onto the focal plane by two cryogenic lenses.  Blackened glint baffling and field stop are used to absorb stray light within the receiver. A 4 K cold stop and oversized warm mirrors minimize warm spill while a co-moving ground screen and forebaffle are used to further minimize ground pickup. 

\begin{figure}[t]
    \centering
    \includegraphics[width=5.4in]{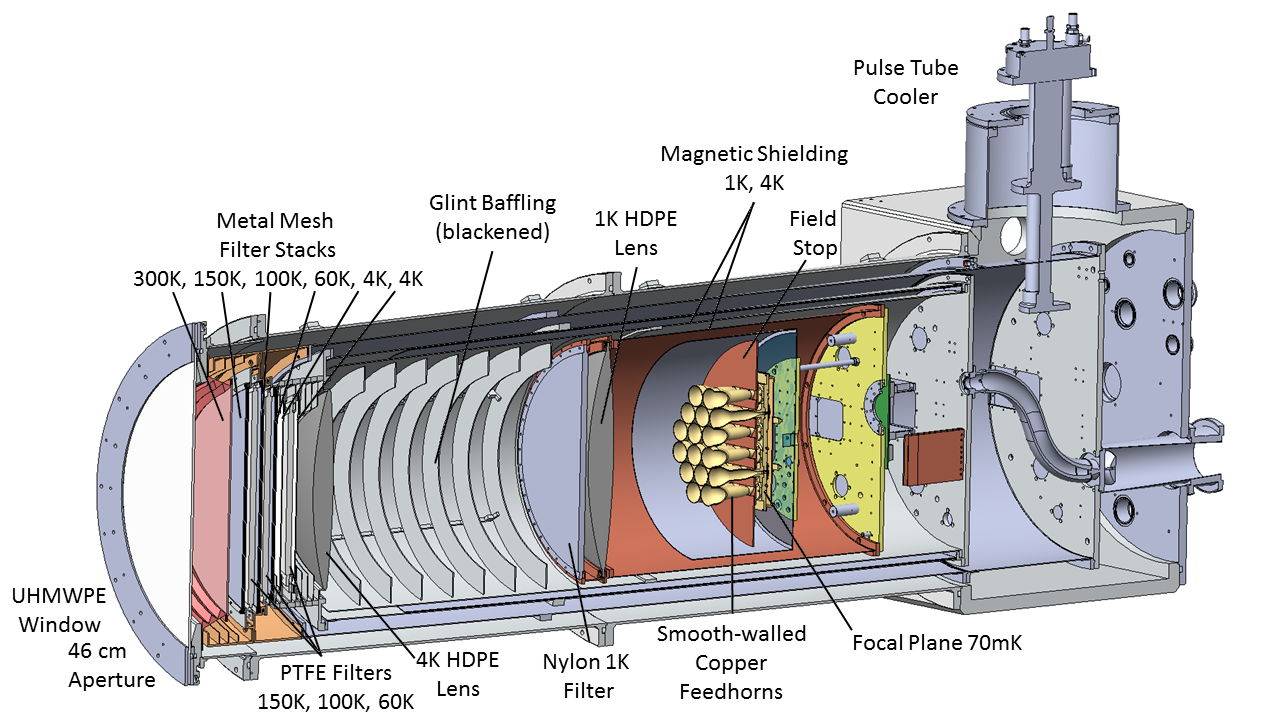}
    \caption{\footnotesize The 38~GHz receiver as it was fielded in early 2016. Vacuum is held over a 46~cm aperture by a 4.8~mm thick ultra-high molecular weight polyethylene window. Six multi-layer stacks of reflective metal-mesh filters interspersed between three polytetrafluoroethylene (PTFE) filters drastically reduce the loading from infrared radiation before the 4~K cold stop. Blackened glint baffling and a blackened field stop absorb stray light while two high-density polyethylene (HDPE) lenses image incident light onto the focal plane. The focal plane sits at 70~mK and is surrounded by two layers of magnetic shielding.}
    \label{fig:qband_receiver}
\end{figure}

The CLASS telescopes use four custom cryostats built by BlueFors Cryogenics.\footnote{BlueFors Cryogenics, Arinatie 10, 00370 Helsinki, Finland, www.bluefors.com} As shown in Figure~\ref{fig:qband_receiver}, the cryostats are pulse tube cooled dilution refrigerators configured with horizontally extended cold stages at 70~mK, 1~K, 4~K and 60~K and a 46 cm aperture.\cite{essinger/etal:2014} The fielded 38~GHz receiver temperatures are stable and at the designed specifications for all telescope orientations.
 
\subsubsection{Lenses}

The 38~GHz and 93~GHz CLASS telescopes each have two high-density polyethylene (HDPE) cryogenic lenses, one at 1~K (Figure \ref{fig:lens_vpm}) and one at 4~K. These lenses are machined at Johns Hopkins University using a Tormach PCNC 1100 Mill\footnote{http://www.tormach.com/}. Each lens was annealed before machining, rough cut into approximately the desired shape, annealed again, and then machined to the final dimensions. The machining errors on all the lens surfaces are less than 100~$\mu$m across the 40~cm diameter and the thicknesses of the lenses are correct to 30~$\mu$m. These errors are well within the tolerances of the optical design.

The HDPE lenses are anti-reflection coated using simulated dielectrics, a technique where sub-wavelength features are cut into the surface of the material to create a layer with lower mean density. By specifically tuning the density of the layer, we optimize the index of refraction to minimize reflection off the surface. The CLASS simulated dielectrics, shown in the left image of Figure \ref{fig:lens_vpm}, use a square array of holes, about a fifth of a wavelength in diameter, cut into the surface of the lenses. The diameter, depth, and spacing of these features are based on electromagnetic simulations created using HFSS\footnote{http://www.ansys.com/Products/Electronics/ANSYS-HFSS} and verified by laboratory measurements.

\subsubsection{Filters}

All CLASS telescopes use a combination of cryogenic reflecting and absorbing filters to reject out of band power and reduce optical loading on the detectors. For the 38~GHz telescope, six multi-layer stacks of reflective metal mesh filters, aluminum square grids on Mylar or polypropylene films, function as low-pass filters and reflect over 75\% of the incident infrared radiation back out the receiver window.\cite{essinger/etal:2014} 

The 38~GHz and 93~GHz telescopes use three polytetrafluoroethylene (PTFE) filters, two on the 60~K cold stage and one on the 4~K cold stage, as well as one nylon filter on the 1~K cold stage to absorb above band frequencies. The 38~GHz PTFE and Nylon filters are anti-reflection coated using simulated dielectrics in a manner similar to the 38~GHz lenses. The 93~GHz PTFE and Nylon filters are anti-reflection coated using heat pressed expanded sheets of porous PTFE. 

\subsubsection{Window}

The 46~cm diameter window for the CLASS 38 GHz telescope is made of 4.8~mm ultra-high molecular weight polyethylene (UHMWP) that is anti-reflection coated with 1.6~mm thick sheet of porous PTFE. Based on the level of bowing observed in the 38~GHz window at 5200~m, the 93~GHz telescopes will have a thinner 3.2~mm UHMWP window that is also anti-reflection coated using porous PTFE.

\begin{figure}[t]
    \centering
    \includegraphics[width=3.1in]{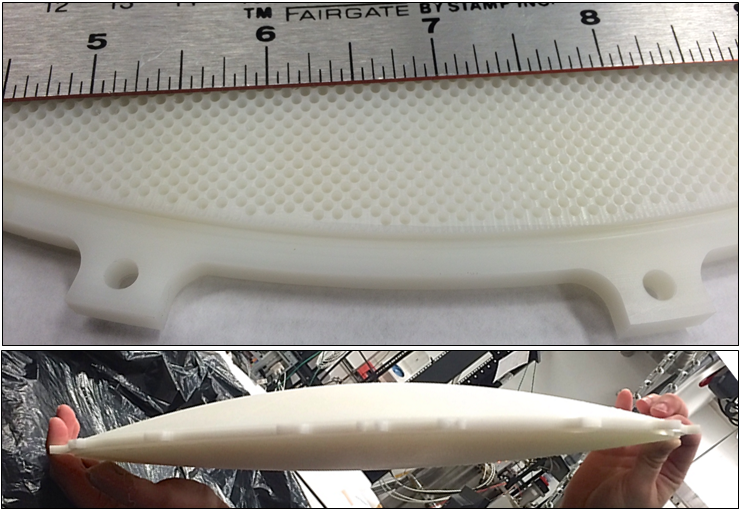}
    \includegraphics[width=2.3in]{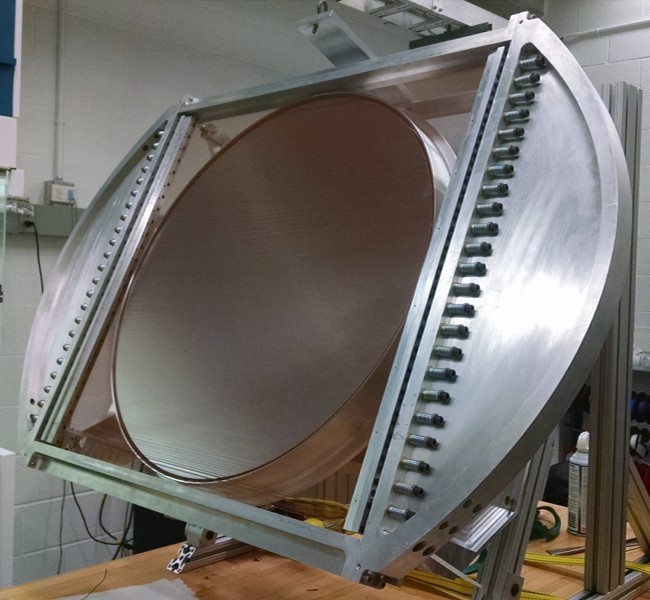}
    \caption{\footnotesize {\bf Left Figure:} (Bottom) The 40 cm diameter 1 Kelvin high density polyethylene lens for the 38 GHz telescope. (Top) A close up of the simulated dielectric anti-reflection coating used for all plastic cryogenic optics in the 38 GHz telescope. {\bf Right Figure:} The variable-delay polarization modulator (VPM) for the CLASS 38 GHz telescope. The wire grid contains over 3000 50$\mu$m copper plated tungsten wires spaced 150$\mu$m apart across a 61 cm clear aperture. A 60 cm flat aluminum honeycomb mirror attached to a mirror transport mechanism sits behind the wire grid and creates a phase delay between polarization states parallel and perpendicular to the wire grid which is modulated at 10 Hz.}
    \label{fig:lens_vpm}
\end{figure}

\subsection{Variable-delay polarization modulator}

The CLASS VPMs consist of a linearly polarizing wire grid in front of a movable flat mirror. The distance between the grid and the movable mirror creates a variable phase delay between polarization states parallel and perpendicular to the wires. The phase delay modulates the incoming Stokes parameters such that Stokes U or Q and V can be fully determined by moving the grid-mirror distance through a quarter of a wavelength. The grid-mirror distance is modulated at 10 Hz, much quicker than the rate of instrument and atmospheric temperature drifts. This rapid modulation, combined with the placement of the VPM as the first optical element in the CLASS telescopes, gives the CLASS telescopes the stability required to measure the large angular scale E- and B-modes across 70\% of the sky. More information on the optical properties of a VPM can be found in Reference \citenum{chuss/etal:2012} and a description of instrument stability provided by the VPM is in Reference \citenum{miller/etal:2016}.

The VPMs for the CLASS telescopes, shown in Figure \ref{fig:lens_vpm}, consist of two main components, a linearly polarizing wire grid and a mirror transport mechanism (MTM). The wire grid uses 50 $\mu$m copper-plated tungsten wires spaced 150 $\mu$m apart across a 61 cm clear aperture. 

The VPM MTM uses cross-bladed rotational flexures in a four-bar-linkage configuration to achieve a large (2.54~mm) mirror throw while maintaining excellent parallelism with respect to the wire grid. A wire flattening ring defines a plane for the wire grid and is used to align the wires parallel to the mirror. For the 38~GHz VPM, the wire plane is parallel to the mirror at a level of 50~$\mu$m across 60 cm. The residual tilt of the VPM mirror is measured using three 0.1~$\mu$m resolution encoders attached to the back of the mirror. Across the 2.54~mm throw, the VPM mirror tilts by less than 10~arcseconds at all telescope orientations. There is also an over-all change in mirror tilt observed as a function of elevation and boresight angle. The average mirror tilt changes by 7~arcseconds as the telescope is moved between 30~degrees and 70~degrees elevation. These tilts are all significantly smaller that the 90~arcmin beams of the 38~GHz CLASS telescope. 

A second reaction canceling axis is built into the MTM and used to cancel the vibrations induced by the mirror motion. Both MTM axes are driven by voice coils, and the positions are read out using Renishaw Atom\footnote{http://www.renishaw.com/en/atom-incremental-encoder-system-with-rtlf-linear-scale--24183} incremental encoders with a resolution of 0.1~$\mu$m. A frequency domain feedback loop maintains the mirror throw amplitude and mean distance from the wire grid within 50~$\mu$m of the set positions while the telescope is scanning in azimuth. 

\section{Conclusions}

We have given an update on the scientific projections and the instrument status for the Cosmology Large Angular Scale Surveyor (CLASS). CLASS will observe the large angular scale ($\geq2^{\circ}$) polarization fluctuations in the cosmic microwave background over 70\% of the sky in four frequency bands (38 GHz, 93 GHz, 145 GHz, 217 GHz). Through these observations, we will make a cosmic variance limited measurement of the optical depth to reionization and characterize the primordial gravitational wave sourced B-modes if the tensor-to-scalar ratio, $r$, is greater than 0.01.

The CLASS 38 GHz telescope was deployed to the CLASS site in the Chilean Atacama desert and achieved first light in May 2016, followed by commissioning activities. Our cryogenics are stable in all operating configurations and our detector loading and saturation powers are within their designed ranges. All systems for the first 93 GHz are in the later stages of fabrication and we will to deploy the two 93~GHz telescopes followed by the dichroic 145/217~GHz telescope over the first two years of the five year CLASS survey.

\section*{Acknowledgements}

We acknowledge the National Science Foundation Division of Astronomical Sciences for their support of CLASS under Grant Numbers 0959349 and 1429236. Detector technologies were developed under previous and continuing NASA awards including NASA grant number NNX14AB76A.  We are also grateful to NASA for their support of civil servants engaged in state-of-the-art detector technologies. K. Harrington is supported by NASA Space Technology Research Fellowship grant number NX14AM49H. T. Essinger-Hileman was supported by an NSF Astronomy and Astrophysics Postdoctoral Fellowship. We further acknowledge the very generous support of Jim and Heather Murren, Matthew Polk, and Michael Bloomberg. CLASS is located in the Parque Astronómica Atacama in northern Chile under the auspices of the Comisión Nacional de Investigación Científica y Tecnológica de Chile (CONICYT).

\bibliography{class}   %>>>> bibliography data in report.bib
\bibliographystyle{spiebib}   %>>>> makes bibtex use spiebib.bst

\end{document}